\documentclass[a4paper]{jpconf}
\usepackage{graphicx}
\begin{document}
\title{Results of dark matter searches with the ANTARES neutrino telescope}

\author{J.D. Zornoza}
\address{IFIC (University of Valencia - CSIC, c/ Catedr\'{a}tico Beltr\'{a}n, 2, Paterna (Valencia) Spain}

\ead{zornoza@ific.uv.es}

\author{C. Toennis}
\address{IFIC (University of Valencia - CSIC, c/ Catedr\'{a}tico Beltr\'{a}n, 2, Paterna (Valencia) Spain}

\ead{ctoennis@ific.uv.es}

\begin{abstract}
Neutrino telescopes have a wide scientific scope. One of their main goals is the detection of dark matter, for which they have specific advantages.  Neutrino telescopes offer the possibility of looking at several kinds of sources, not all of them available to other indirect searches. In this work we provide an overview of the results obtained by the ANTARES neutrino telescope, which has been taking data for almost ten years. One of the most interesting ones is the Sun, since a detection of high energy neutrinos from it would be a very clean indication of dark matter, given that no significant astrophysical backgrounds are expected, contrary to other indirect searches. Moreover, the limits from neutrino telescopes for spin-dependent cross section are the most restrictive ones. Another interesting source is the Galactic Centre, for which ANTARES has a better visibility than IceCube, due to its geographical location. This search gives limits on the annihilation cross section. Other dark matter searches carried out in ANTARES include the Earth and dwarf galaxies.
\end{abstract}

\section{Introduction}
The nature of dark matter (DM) is one of the most relevant topics in Physics today. Even if the list of experimental evidence is large and variate, we still do not know what is made of. Given the fact that many of the properties of the particle (or particles) forming the dark matter in the Universe are unknown, it is important to use different approaches. As it will be explained in the following, neutrino telescopes have specific advantages and can explore models not accessible to other searches.

\section{The ANTARES neutrino detector}

The ANTARES neutrino telescope~\cite{antares} is installed in the Mediterranean Sea, about 40~km off the French coast, near Toulon, at a depth of 2500~m. It is made of 885 photomultimpliers (PMTs) housed in glass spheres, forming the so-called optical modules (OMs). The OMs are installed in 12 lines anchored at the sea bottom and kept vertical by buoys. 

The operation principle is based on the detection of the Cherenkov light induced in water by the leptons produced after the weak interaction of neutrinos in the surroundings of the detector. The main channel, given its best angular resolution, are the muon tracks  from the CC interactions of muon neutrinos. This is the only channel used in this analysis. For a review of recent results of the ANTARES experiment see~\cite{antoine}.

\section{Dark matter detection}

Neutrino telescopes are an important and complementary experimental approach to detect dark matter. This technique offers some unique advantages. The search for dark matter with neutrino telescopes is based on the idea that dark matter particles would self-annihilate. In most models, direct neutrino production is not expected, but the decay of the secondary particles produced in the annihilations would yield neutrinos that can reach the detector. Many models assume that the DM particle is a weakly interacting massive particle (WIMP).

There are several sources interesting for dark matter searches. In the case of the Sun or the Earth, the DM particles are scattered off with the nuclei in these astrophysical bodies and lose energy, so they can become gravitationally bound, accumulate and self-annihilate. In this case, the results on the neutrino flux can be related to the WIMP-nucleon scattering cross section, analogously to direct searches. For this, equilibrium between capture and annihilation must be assumed, which is a reasonable assumption for the Sun, but not in the case of the Earth. 

The second kind of searches is the case of astrophysical bodies where dark matter gathers by gravitational accumulation along the process of structure formation. The candidate sources are the Galactic Centre, dwarf galaxies and galaxy clusters. For these, the DM annihilation cross section can be measured or constrained. Given its proximity, the best results with neutrino telescopes are obtained for the Galactic Centre.

\section{Sun}

The Sun is one of the most interesting sources to be looked at with neutrino telescopes. The main advantage of this analysis is that it is free from astrophysical uncertainties, contrary to cases like searches with gamma-rays or cosmic rays. For the latter, recent hints are very difficult to interpret as a genuine dark matter signal since other explanations (pulsars, supernova remnants, etc.) are also plausible. For neutrino telescopes, on the contrary, high energy neutrinos are only expected from dark matter scenarios. The background comes only from atmospheric muons and atmospheric neutrinos, the rate of which can be robustly estimated from data. A contamination from high energy cosmic rays interacting in the Sun's corona and producing high energy neutrinos is negligible.

\begin{figure}[h]
	\includegraphics[width=20pc]{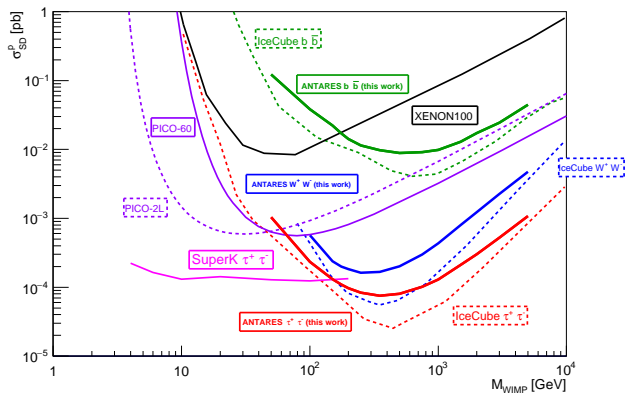}\hspace{2pc}%
	\begin{minipage}[b]{14pc}\caption{\label{sdcs}Limits set by ANTARES on the WIMP-p scattering cross section (spin-dependent) as a function of the WIMP mass, compared to other experiments.}
	\end{minipage}
\end{figure}

The results of the search for dark matter in the Sun using ANTARES data of 2007-2012 are presented in Figure~\ref{sdcs}. It shows the limits on the spin dependent WIMP-proton cross section. For this case neutrino telescopes (ANTARES, SuperKamiokande, IceCube) obtain limits which are better than any direct search. For the spin-independent cross section this is not the case.

\section{Galactic Centre}

The Galactic Centre is also a major candidate for DM searches with neutrino telescopes. As mentioned above the results are interpreted in terms of annihilation cross section, so in some sense they are complementary to the analysis of the Sun. Moreover, in this case the high energy component of the neutrino spectrum arrives unaffected to the detector, while in the Sun part of these neutrinos are absorbed.

\begin{figure}[h]
	\begin{minipage}{14pc}
		\includegraphics[width=20pc]{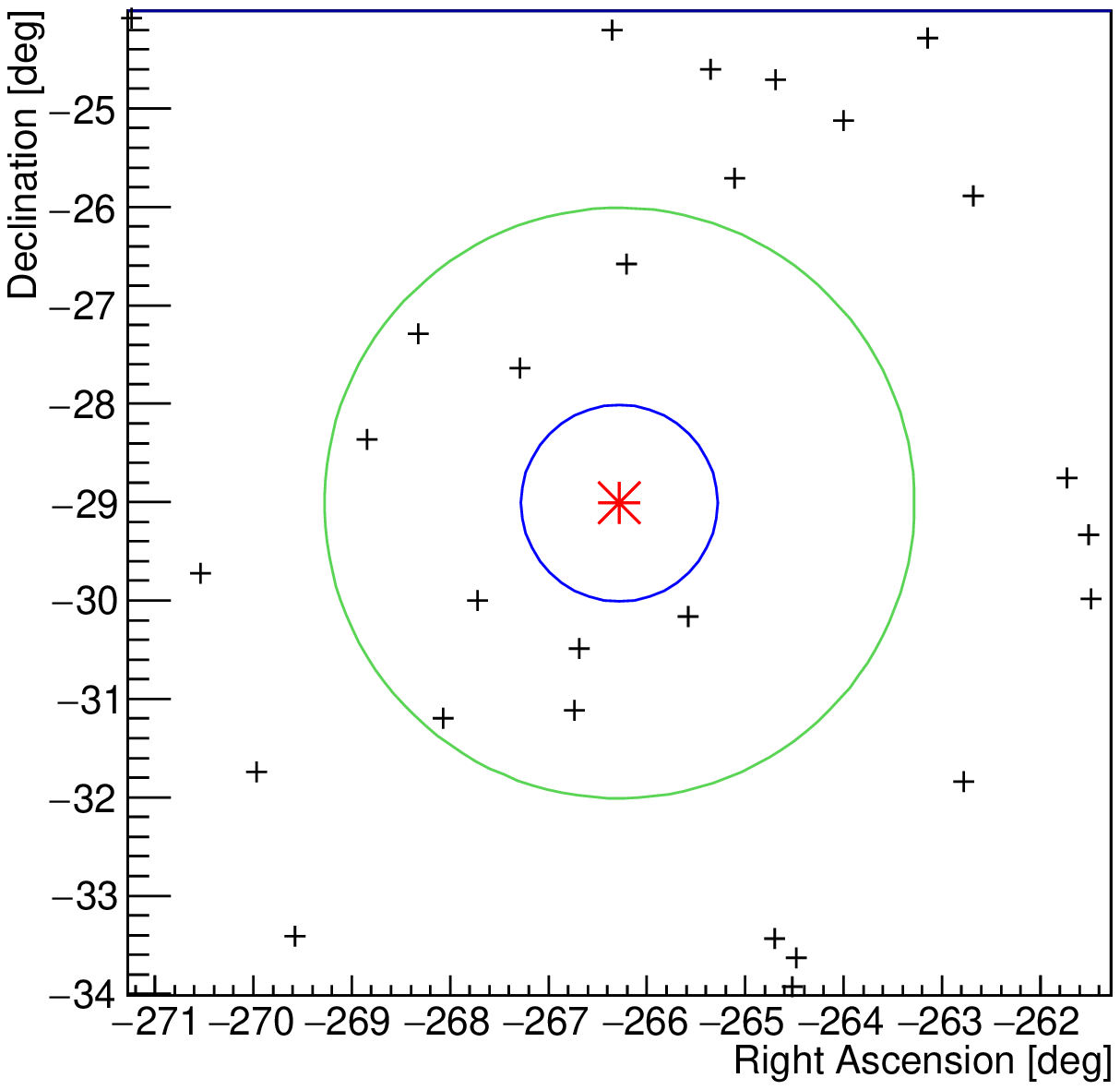}
	\end{minipage}\hspace{2pc}%
	\begin{minipage}{14pc}
		\includegraphics[width=20pc]{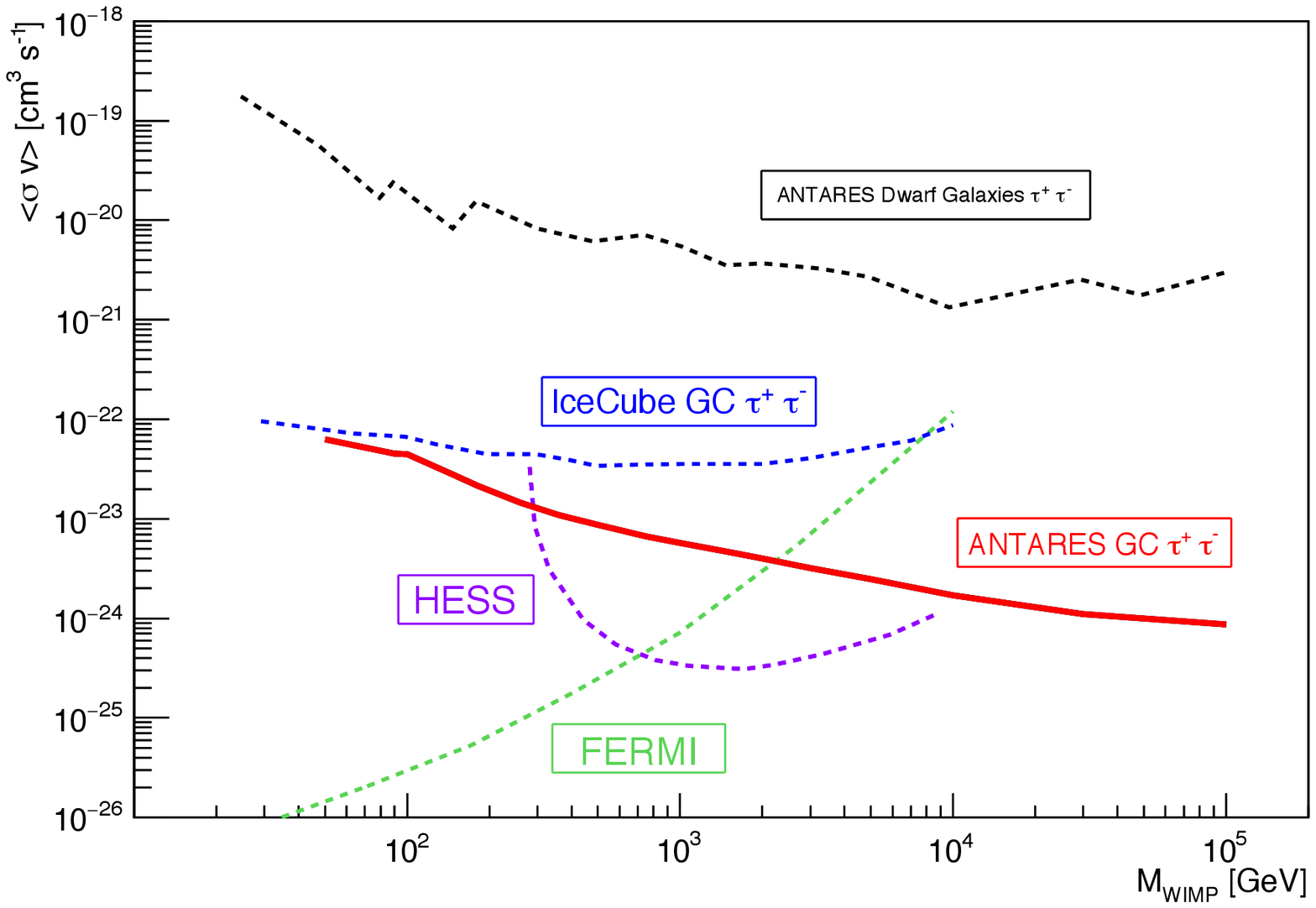}
	\end{minipage} 
		\caption{\label{gclim}Left: Distribution of events in the sky around the Galactic Centre in the ANTARES data of 2007-2013. Right: Limits set by ANTARES on the WIMP annihilation cross section as a function of the WIMP mass, compared to other experimental searches, for the $\tau$-$\tau$ channel and the NFW halo profile.}

	\label{gc}
\end{figure}

Figure~\ref{gc} shows the results of the search for neutrinos in the direction of the Galactic Centre. These limits are better than those of its main competitor, IceCube, since the visibility of the Galactic Centre is better for ANTARES due to their respective geographical locations. For masses above 10 TeV, ANTARES provides the best limits in the world.

\section{Conclusions}

We have presented the results of several dark matter searches done with ANTARES. For the Sun, the results of neutrino telescopes, including those of ANTARES (with data of 2007-2012), offer the best limits for spin-dependent WIMP-p cross section and a case free of astrophysical uncertainties. For the Galactic Centre, ANTARES (with data of 2007-2013) set the best limits among neutrino telescopes, which are also best of any technique for WIMP masses above 10 TeV.

\section*{Acknowledgments}

The authors thank the support of Plan Estatal de Investigaci\'{o}n (refs. FPA2015-65150-C3-1-P, -2-P and -3-P, (MINECO/FEDER)), Severo Ochoa Centre of Excellence and MultiDark Consolider (MINECO), and Prometeo and Grisol\'{i}a programs (Generalitat Valenciana), Spain.


\section*{References}
\begin{thebibliography}{9}
\bibitem{antares}M. Ageron  {\it et al.}  (ANTARES collaboration), Nucl. Instrum. Meth. A 656, 11-38 (2011)
\bibitem{antoine} High-energy neutrino searches in the Mediterranean Sea: probing the Universe with ANTARES and KM3NeT/ARCA, A. Kouchner on behalf of the ANTARES Collaboration, proc. of this conference
\bibitem{sun_paper} 
 S. Adri\'{a}n et al., ANTARES Collaboartion, Physics Letters B, Volume 759 (2016)
\end{thebibliography}
\end{document}